\pdfoutput=1

\documentclass[reprint,
superscriptaddress,
 amsmath,amssymb,
 aps,
longbibliography,
prl
]{revtex4-2}

\usepackage{xparse}

\usepackage{appendix}
\usepackage{graphicx}
\graphicspath{ {./Figures/} }

\usepackage{dcolumn}
\usepackage{bm}
\usepackage{hyperref}
\usepackage{xcolor} 

\usepackage[noabbrev]{cleveref}

\usepackage{comment}
\usepackage{dsfont}

\newcommand{\ket}[1]{\ensuremath{\left| #1 \right>}}

\usepackage{slashed}
\renewcommand{\vec}[1]{\mathbf{#1}}

\usepackage{bm}
\newcommand{\vect}[1]{{\boldsymbol{\mathbf{#1}}}}

\usepackage{scalerel}

\newcommand{\tri}{\triangledown}

\newcommand{\sgn}{\text{sgn}}
\usepackage{cancel}
\usepackage{nicefrac}
\usepackage{xfrac}
\usepackage{textcomp}
\usepackage{romannum}
\usepackage{mathtools} 

\usepackage{bbold}

\begin{document}
\title{Zero Flux Localization: Magic Revealed}

\author{Alireza Parhizkar}
    \affiliation{Joint Quantum Institute, Department of Physics, University of Maryland, College Park 20742}

\author{Victor Galitski}
    \affiliation{Joint Quantum Institute, Department of Physics, University of Maryland, College Park 20742}

\begin{abstract}
Flat bands correspond to the spatial localization of a quantum particle moving in a field with discrete or continuous translational invariance. The canonical example is the flat Landau levels in a homogeneous magnetic field. Several significant problems---including flat bands in moir\'e structures---are related to the problem of a particle moving in an inhomogeneous magnetic field with zero total flux. We demonstrate that while perfectly flat bands in such cases are impossible, the introduction of a ``non-Abelian component''---a spin field with zero total curvature---can lead to perfect localization. Several exactly solvable models are constructed: (i)~a half-space up/down field with a sharp 1D boundary; (ii)~an alternating up/down field periodic in one direction on a cylinder; and (iii)~a doubly periodic alternating field on a torus. The exact solution on the torus is expressed in terms of elliptic functions. It is shown that flat bands are only possible for certain magic values of the field corresponding to a quantized flux through an individual tile.  These exact solutions clarify the simple structure underlying flat bands in moir\'e materials and provide a springboard for constructing a novel class of fractional quantum Hall states.

\end{abstract}


\maketitle

Electrons in a homogeneous background magnetic field form flat bands -- energy levels that are independent of the momentum of the electron $\vec{k} \equiv (k_x,k_y)$. This Landau quantization implies that the electrons are spatially localized,  since $\partial E / \partial \vec{k} = 0$. In an inhomogeneous magnetic field, the momentum  is no longer a good quantum number since the translational invariance is lost. However, one can still ask whether highly degenerate dispersionless energy levels exist. It has been known that if the total flux through the system, $\Phi = N \Phi_0$ with $\Phi_0$ being the flux quantum, is nonzero,  at least $N$ modes with the same energy exist~\cite{Aharonov,ParLocalizing} -- a flat band.  

But what if the total flux is zero? This is the principal question explored in this Letter. It is of relevance to time-reversal symmetric systems such as bilayer graphene~\cite{MacDonald,balents,EmEnSc,Exp2018unconventional,Exp2018correlated} and TMDs~\cite{TMDSpin,TMDPseudo,TMDStrain,TMDEffective,TMDCrepel,TopCrepel} where a continuous model can be written in terms of electrons coupled to effective background gauge fields~\cite{NonAbelianGauge} and where  flat bands are known to emerge for certain field configurations~\cite{OriginVishwanath,TopolCrit}. The goal of this work is to elucidate the origin of such zero-flux localization by exploring simple  models, which retain the key mathematical ingredients of relevance to the aforementioned physical systems but allow an exact analytical solution.  This provides intuitive examples, which reveal the underlying mechanism of emerging flat bands in a gauge field with zero net curvature.

The motion of a relativistic electron with linear dispersion in the $x$–$y$ plane in the presence of a magnetic field is described by the Hamiltonian,
\begin{equation} \label{LinH}
    h \equiv i \vect{\sigma} \cdot \left(\vect{\nabla} - ie\vect{A}\right) \, ,
\end{equation}
where $\vec{A}$ is the electromagnetic gauge field and $\vect{\sigma}\equiv (\sigma^x,\sigma^y)$ is the vector of Pauli matrices. Squaring $h$ gives rise to a Hamiltonian that describes an electron with quadratic dispersion in the same magnetic field,
\begin{equation} \label{QuadH}
    h^2  \equiv - \left(\vect{\nabla} - ie\vect{A} \right)^2 \mathbb{1} - eB\sigma^z \equiv H \mathbb{1} - eB\sigma^z \, ,
\end{equation}
where $B = |\nabla \times \vec{A}|$ is the strength of the magnetic field in the $\hat{\vec{z}}$ direction and $\mathbb{1}$ is the $2 \times 2$ identity matrix. Apart from the Zeeman term, $eB\sigma^z$,  $H$  contains a massive electron with mass $m=1/2$ in natural units where $\hbar=c=1$~\footnote{We will also absorb $e$ into the definition of $\vec{A}$, effectively setting it to unity. However, we will still occasionally display the symbol for clarity.}. For a constant background magnetic field, the classical trajectory of such an electron is a circle with the cyclotron radius $r=mv/eB$ where $v$ is the velocity of the particle. Semiclassically, the circumference of the circle must be an integer multiple of the wavelength of the particle, $\lambda = \hbar/mv$, so that the particle goes back to itself after a cycle. This quantizes the flux enclosed by the cyclotron orbit to an integer number of the flux quanta $\Phi_0 = \frac{\hbar}{2 \pi e}$. More precisely, the Aharonov-Bohm phase of the particle traversing the cyclotron orbit $\oint \vec{A} \cdot \vect{d\ell} = 2 \pi N$ must be an integer multiple of $2\pi$. 

\begin{figure}
\includegraphics[width=\linewidth]{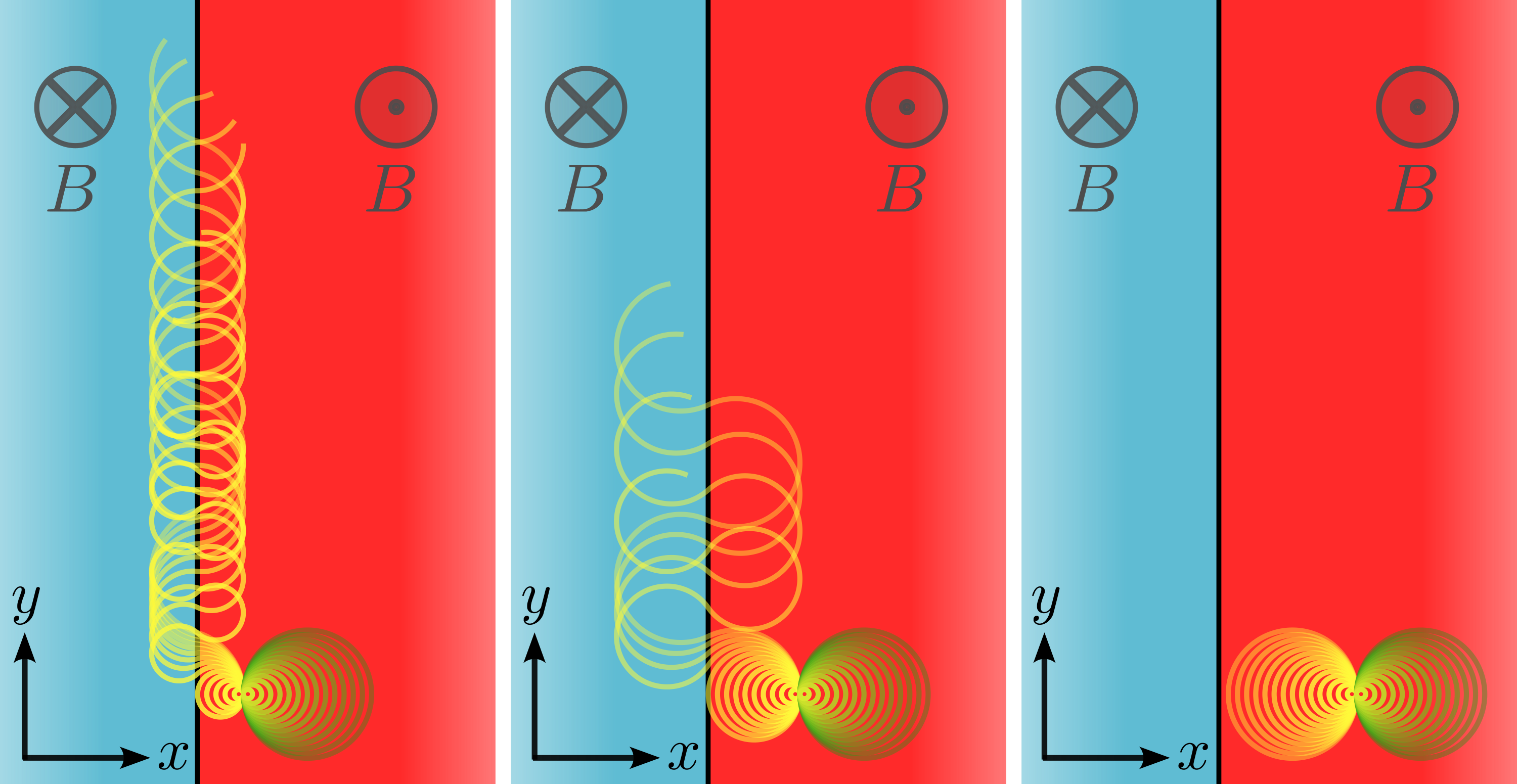}
\centering
\caption{Classical trajectories of a charged particle in a magnetic field that flips direction at $x=0$. Each trajectory corresponds to a different initial velocity along $\hat{y}$. The blue and red regions indicate areas where $B<0$ and $B>0$, respectively. The particle's initial position is progressively closer to the boundary at $x=0$ from right to left. When the magnetic field is homogeneous, the probability of the particle returning to the same position (return probability) is one. However, in this field configuration, unbounded trajectories always exist. Using a Feynman path integral argument, this implies that localization is impossible in the quantum case, as there is always a finite amplitude for a transition to a state that escapes to infinity.}
\label{fig:Traj}
\end{figure}

This picture breaks down if instead of a constant magnetic field, we have an inhomogeneous magnetic field with zero total flux through the system. Consider the configuration where the translational symmetry is broken in the $\hat{x}$ direction, with a downward magnetic field in the $x < 0$ region of the $x$--$y$ plane and an upward magnetic field in the $x > 0$ region, both with the same strength: $\vec{B} = \sgn(x)B \hat{\vec{z}}$. Classically, particles with sufficiently large velocity and cyclotron radius can reach the boundary at $x = 0$. However, upon crossing the dividing line, they must reverse the handedness of their motion; if they were rotating clockwise, they will now rotate counterclockwise with the same radius, and vice versa. This reversal creates a net motion along the $\hat{y}$ direction at the boundary, disrupting localization at the classical level. For the chosen directions of the magnetic field, and for a positively charged particle, the only physical trajectory near $x = 0$ is upward along $\hat{y}$, as shown in Fig.~\ref{fig:Traj}.

\begin{figure}
\includegraphics[width=\linewidth]{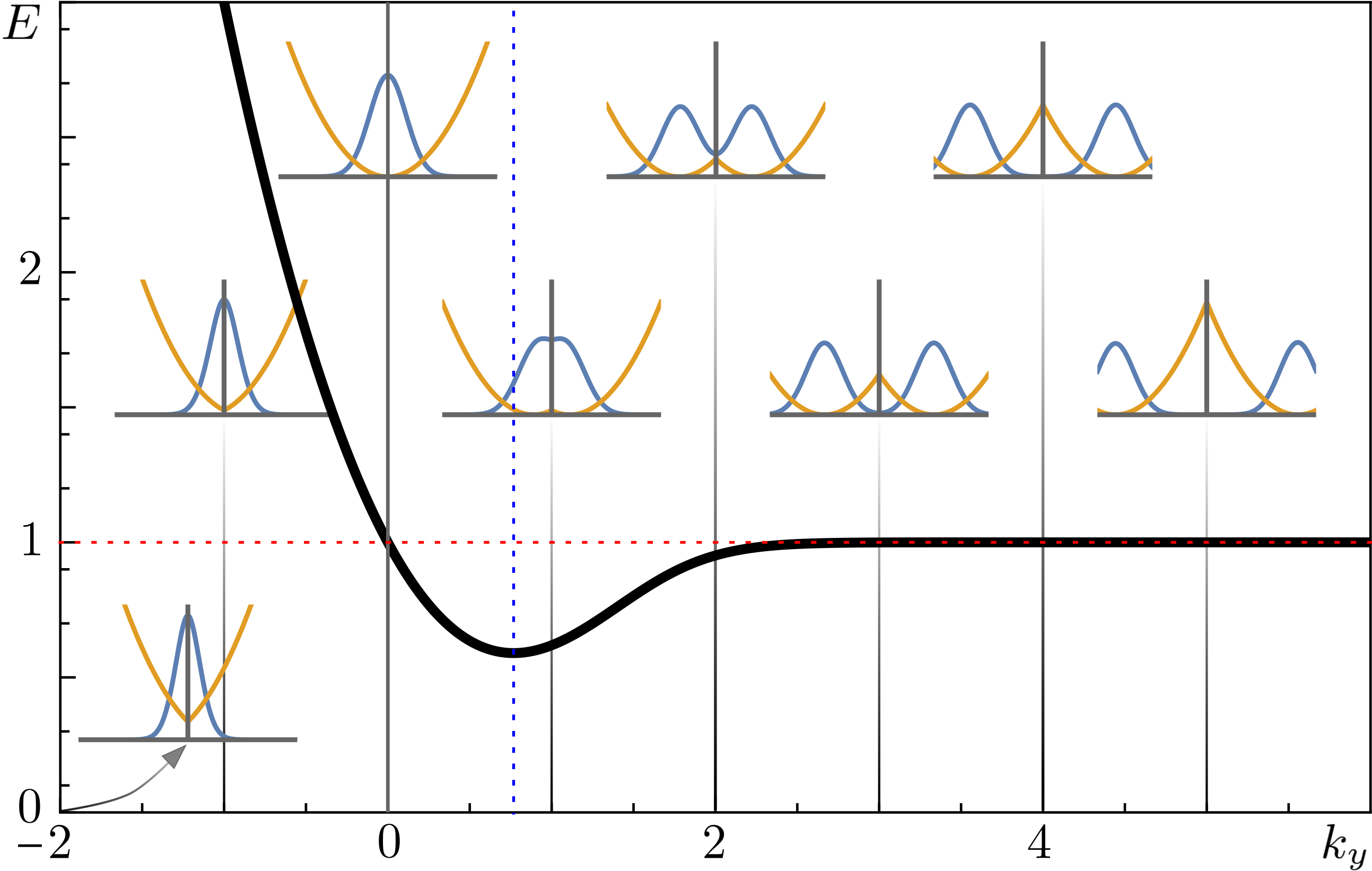}
\centering
\caption{Plotted in black is the numerical solution of the energy spectrum, $E(k_y)$, of the Schr{\"o}dinger operator~(\ref{SE})  as a function of $k_y$ in units of $2\pi/\ell_B \equiv 2\pi\sqrt{eB}$. The smaller plots depict the one dimensional potential, $V_{k_y}(x) =  (eB|x| - k_y )^2$ in yellow and the corresponding ground state wave-function in blue (the plots are positioned along the $x$-axis according to the value of $k_y$; their vertical position carries no significance; arbitrary units are used.) Note that the group velocity $E'(k_y)$ is negative before the minimum of the potential at $k_y\approx 0.77$ and it becomes positive afterwards. At large $k_y$, the band is almost flat, which corresponds to two harmonic oscillator potentials far away from each other.}
\label{fig:Eigen}
\end{figure}

The quantum picture has more to offer. To maintain translational symmetry along $\hat{y}$, we choose the gauge $\vec{A}(x,y) = B|x|\hat{y}$ which yields $\vec{B}=\vect{\nabla}\times \vec{A} = \sgn(x)B \hat{\vec{z}}$. The corresponding Schr{\"o}dinger equation reads,
\begin{align}
\label{SE}
    &H \psi(x,y)= H e^{ik_y y} \psi(x) = E(k_y) e^{ik_y y}\psi(x) \\
    &\text{while} \quad \left[-\partial_x^2 + (eB|x| - k_y )^2 \right]\psi(x) = E(k_y) \psi(x) \, . \nonumber 
\end{align}
The potential appearing in the one-dimensional eigenvalue problem above is depicted in Fig.~\ref{fig:Eigen}. Substituting $|x| \rightarrow x$ gives us the usual Landau quantization back where the shape of the one-dimensional potential remains unchanged by varying $k_y$, making the eigenvalues independent of $k_y$. In the current scenario, however, the shape of the potential depends on $k_y$ and hence also the eigenvalues of $H$. For simplicity let us focus on the ground state energy. At $k_y=0$ we recover the simple harmonic oscillator potential, where the ground state energy is $eB$. Decreasing $k_y$ below zero raises the minimum of the potential, thus raising the ground state energy as well. So $\partial E/\partial k_y|_{k_y=0} <0 $.
When $k_y$ is large and positive, $k_y \gg 0$, we have two well-separated potential wells and the problem is practically reduced to that of two isolated harmonic oscillators which again have the ground state energy of $eB$. From the points above one concludes that there is a local minimum of ground state energy between $k_y=0$ and $k_y \rightarrow \infty$. (Fig.~\ref{fig:Eigen})

The classical prohibition of downward trajectories is reframed on the quantum side in having no low energy states for $k_y<0$ due to the prohibiting shape of the one-dimensional potential with ever rising minima as $k_y$ decreases (Fig.~\ref{fig:Eigen}). However wavenumber (or momentum) alone does not represent propagation, instead that is determined by the group velocity $\partial E/\partial k_y$ which in the current setup shows a different behaviour depicted in Fig.~\ref{fig:Eigen}: Even though $k_y<0$ are forbidden,
there still exists downward propagation.

The massive electron is not localized in the magnetic field above. Is there a way to restore the lost localization? The obstruction to localization and its resolution are revealed most clearly in the relativistic case. Let us go back to the Dirac electron,~Eq.~\eqref{LinH}, with the gauge field $\vec{A}(x,y) = B|x|\hat{y}$, and consider its zero modes $h\Psi \equiv h[\psi^+,\psi^-]^T=0$. The corresponding Dirac equation 
\begin{align}
    \!\! \left[
    \begin{array}{cc}
     0 & \!\!\!\!\!\!\!\!\partial_x - iA_x - i\partial_y  - A_y  \!\\
     \!\partial_x - iA_x + i\partial_y  + A_y & \!\!\!\!\!\!\!\!0
    \end{array}
    \right]\!\!
    \left[ \begin{array}{cc}
     \psi^+ \\
     \psi^-   
    \end{array}  \right] \!\!  = \!   0 \, ,
\end{align}
splits into two decoupled equations,
\begin{equation}
    \left[ (\partial_x \mp A_y) \mp i (\partial_y \pm A_x) \right] \psi^\mp=0 \, .
\end{equation}
They  have the following solutions~\cite{Aharonov,ParLocalizing},
\begin{equation} \label{Solutions}
    \psi^\pm = f^\pm (x \pm iy) e^{\pm\phi} \, ,
\end{equation}
for all $f^\pm(x\pm iy)$, while $\phi$ is defined to solve    
\begin{equation} \label{Poisson}
   \partial_x \phi = -A_y \quad \text{and} \quad \partial_y \phi = A_x \quad (\Rightarrow \vect{\nabla}^2\phi = -B) \, .
\end{equation}
which always has a solution for any given magnetic field. According to Eq.~\eqref{Solutions} there are infinitely many solutions to $h\Psi=0$. However, the physical zero-modes are only those solutions which are normalizable: $\int \! d^2x \Psi^\dagger \Psi =1$. The number of physical zero-modes is determined by the total magnetic flux threading the system~\cite{ParLocalizing,Aharonov}. For a constant homogeneous magnetic field $\vect{\nabla}^2\phi=-B$ is satisfied by $\phi= -\frac{B}{2}x^2$. Without loss of generality we have disregarded the $y$ dependent solutions of $\phi$.
For $B>0$ this gives a bounded $\psi^+$ according to Eq.~\eqref{Solutions} while $\psi^-$ will be unbounded. So the $\psi^-$ component of $\Psi$ is troublesome and it should be eliminated by setting $f^-(x-iy)=0$, otherwise there will be no normalizable zero-mode $\Psi$. Thus, for $\vec{B}=B \hat{\vec{z}}$,  we have the following normalizable zero-modes,
\begin{equation} 
    h\Psi^+ \equiv h \left[ \begin{array}{cc}
     \psi^+ \\
     0   
    \end{array}  \right]
    = h \left[ \begin{array}{cc}
     f^+ (x+iy) e^{+\phi}\\
     0   
    \end{array}  \right] 
    = 0 \, ,
    \label{ZeroModeBP}
\end{equation}
while for $\vec{B}=-B \hat{\vec{z}}$ only the wavefunctions below are normalizable and hence acceptable as zero-modes,
\begin{equation} 
    h\Psi^- \equiv h \left[ \begin{array}{cc}
     0 \\
     \psi^-   
    \end{array}  \right]
    = h \left[ \begin{array}{cc}
     0 \\
     f^- (x-iy) e^{+\phi}   
    \end{array}  \right] 
     = 0 \, .
    \label{ZeroModeBM}
\end{equation}
Note that consequently neither $\Psi^+$ nor $\Psi^-$ are normalizable when the zero-flux background magnetic field, $\vec{B}=\sgn(x)B \hat{\vec{z}}$, is applied.

For $\vec{B}=\sgn(x)B \hat{\vec{z}}$, therefore, the zero-modes can only be of the form
\begin{equation}
    h\Psi = h \left[ \begin{array}{cc}
     \psi^+ \, \theta(+x) \\
     \psi^- \, \theta(-x)  
    \end{array}  \right]
     = 0 \, ,
    \label{ZeroModeBPM}
\end{equation}
with $\theta(x)$ being the Heaviside step function. This ansatz, $\Psi$, is indeed normalizable, but it does not solve the Dirac equation~\eqref{LinH}, because it is not continuous: An spatial integration of $h\Psi$ over an infinitesimal range $[-\epsilon,\epsilon]$ spits out $\Psi |^{+\epsilon}_{-\epsilon}$ from the derivative part and zero from the potential part. So here the solution must be continuous whereas the only candidate, Eq.~\eqref{ZeroModeBPM}, jumps across the diving line at $x=0$,
\begin{equation}
    \Psi(+\epsilon,y)-\Psi(-\epsilon,y) = 
    \left[ \begin{array}{cc}
     f^+ (iy) e^{+\phi(0,y)}\\
     -f^- (iy) e^{+\phi(0,y)}   
    \end{array}  \right] \, .
\end{equation}
No zero-mode exists.

However, one can rectify this problem by introducing a singular potential on the dividing line. One way to do this is by $h \rightarrow h - \sigma^y \delta(x)$ or in other words by adding an extra gauge field,
\begin{equation}\label{AbelianR}
    h \rightarrow \bar{h} = i\vect{\sigma}\cdot (\vect{\nabla} - i \vect{A} + \vect{S}\sigma^z) \quad \text{with} \quad \vect{S} = (2\delta(x), 0) \, ,
\end{equation}
where the presence of $\sigma^z$ indicates that the new field is a spin (or chiral) field.
With this modification the value of $\Psi$ on the dividing line will now remove the discontinuity problem and we will have, according to Eqs.~\eqref{ZeroModeBP} and \eqref{ZeroModeBM}, infinitely many zero modes. A simple choice of $f^\pm$ such as $f^\pm(x\pm iy)=i^n (x\pm iy)^n$ sets the two functions equal at $x=0$. Although the total flux is zero, $\int d^2x \vec{B}(x)=0$, after introducing $\vec{S}$ the number of normalizable zero modes, $N$, depends on the total absolute flux $N=\int d^2x |\vec{B}(x)|/\Phi_0$ with $\Phi_0$ being the flux quantum~\cite{ParLocalizing,TopolCrit}. Note that there is no flux attributed to the $\vec{S}$ field so the whole system remains flux-less.

The singular nature of the spin field is not important in general. The jump in the wave-function can be smeared out. The simplest example is when we substitute the dividing line by a thick ribbon where the magnetic field, $\vec{B}=\vect{\nabla}\times \vec{A}$, is absent. Let the span of the ribbon be $-\ell <x<\ell$. The electron is governed by the same hamiltonian, $h$, but by setting $\vec{A}=[-\theta(-x-\ell)+\theta(x-\ell)]Bx \hat{\vec{y}}$ instead. We now know that for $x<-\ell$ ($x>\ell$) the wave-function needs to be spin down (up). So for a smooth transition we would want the spin to gradually change as we go from $x=-\ell$ to $x=\ell$ across the ribbon. We can write the wave-function  as follows
\begin{align}
    \tilde\Psi &=
    e^{-\frac{B}{2}(x^2-\ell^2)}
    \left[ \begin{array}{cc}
     0\\
     f^- (x-iy)   
    \end{array}  \right] \theta(-x-\ell)  \nonumber \\
    &+
    \frac{1}{2}\left(1+ \frac{x}{\ell}\sigma^z\right)
    \left[ \begin{array}{cc}
     f^+ (x+iy)\\
     f^- (x-iy)    
    \end{array}  \right] \theta(\ell+x)\theta(\ell-x) \nonumber \\
    &+
    e^{-\frac{B}{2}(x^2-\ell^2)}
    \left[ \begin{array}{cc}
     f^+ (x+iy)\\
     0   
    \end{array}  \right] \theta(x-\ell) \, , \label{Smeared}
\end{align}
while we also add the zero-flux rectifying field $\vec{S} =\hat{\vec{x}}\theta(\ell+x)\theta(\ell-x) \left(1+ \frac{x}{\ell}\sigma^z\right)/2\ell(x^2-\ell^2)$,
to the hamiltonian. During the middle stage in the second line above, when the electron is passing through the ribbon from left to right, the spin smoothly changes from down to up. In order for $\vec{S}$ to have a smooth behaviour at $x=\pm \ell$ the ribbon needs to extend over to magnetic regions. This in turn alters the flux of each region. Having shown that smeared out solutions exist, we are going to carry on with our simple example for the rest of the paper avoiding unnecessary complications.

\begin{figure}
\includegraphics[width=\linewidth]{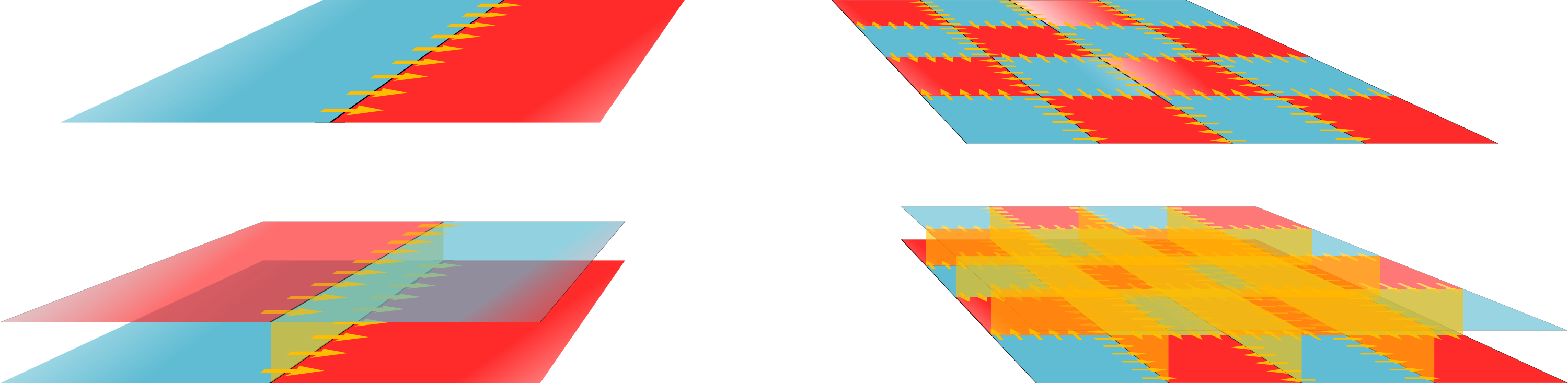}
\centering
\caption{On the left, the simple example with one dividing line turned into a bilayer system with non-Abelian fields. On the right, the simple example is promoted to that of a tiled up surface with multiple dividing lines along with its bilayer (and non-Abelian) counterpart.}
\label{fig:Bilayer}
\end{figure}

Let us develop our simple example into a bilayer system, where on the top layer the electrons are coupled to $\vec{A}_\uparrow (x,y) = B|x|\hat{\vec{y}}$ while on the bottom layer they are coupled to its negative counterpart $\vec{A}_\downarrow (x,y) = -B|x|\hat{\vec{y}}$. The hamiltonian of this model is given by,
\begin{equation}
    h_b = \gamma^0 \vect{\gamma}\cdot \left(\vect{\nabla} - i  \vec{A} \gamma_5 \right) \, ,
\end{equation}
with $\vect{\gamma}\equiv(\gamma^1,\gamma^2)$ and $\gamma$s being the gamma matrices.
In the absence of any other field, following Eqs.~\eqref{ZeroModeBM} to~\eqref{ZeroModeBPM}, the same problematic solution must be given by,
\begin{equation} \label{SolBi}
    \Psi_b = \left[ \psi^+_\uparrow\theta(+x), \psi^-_\uparrow\theta(-x),\psi^+_\downarrow\theta(-x),\psi^-_\downarrow\theta(+x) \right]^T \, ,
\end{equation}
and the discontinuity at $x=0$ with,
\begin{equation} \label{JumpBi}
    \Psi_b(+\epsilon,y)-\Psi_b(-\epsilon,y) = e^{+\phi(0,y)}\left[f^+_\uparrow,-f^-_\uparrow,-f^+_\downarrow,f^-_\downarrow\right]^T \!\! \Big|_{iy} .
\end{equation}
Since now we have a larger degree of freedom there exist more than one way to rectify this problem. One is to set $f^\pm_\downarrow=0$ and then introduce the same field as in Eq.~\eqref{AbelianR} to patch the first and second components of $\Psi$, i.e., intralayer patching. A more nontrivial way is to patch the first and third components instead, i.e., interlayer patching \cite{Bernevig_2006} (See Fig.~\ref{fig:Bilayer}). This is carried out by introducing a spin field proportional to $\gamma_3$ in the Hamiltonian,
\begin{equation} \label{NonAbelianR}
    h_b \rightarrow \bar{h}_b = \gamma^0 \vect{\gamma}\cdot \left(\vect{\nabla} - i \vec{A} \gamma_5 - i\vec{S}i\gamma_3 \right) \, ,
\end{equation}
while the spin vector field $\vec{S}$ is defined as before in Eq.~\eqref{AbelianR}. The interesting feature here is that in order to preserve the flat band we have patched solutions in a non-Abelian way, namely, that the overall gauge field, $\vec{A}\gamma_5 + \vec{S}i\gamma_3$, does not commute with itself at every point. This exactly resembles the structure of the mutually deformed (twisted or strained) bilayer graphene. Compare the action of the above theory,
\begin{equation} \label{BiS}
    S = \int\! d^3x \, \bar\psi i\gamma^\mu (\partial_\mu - i  A_\mu\gamma_5 - i S_\mu i\gamma_3) \psi \, ,
\end{equation}
with that of bilayer graphene, e.g. in Refs.~\cite{TopolCrit,ParLocalizing}.

\begin{figure}
\includegraphics[width=\linewidth]{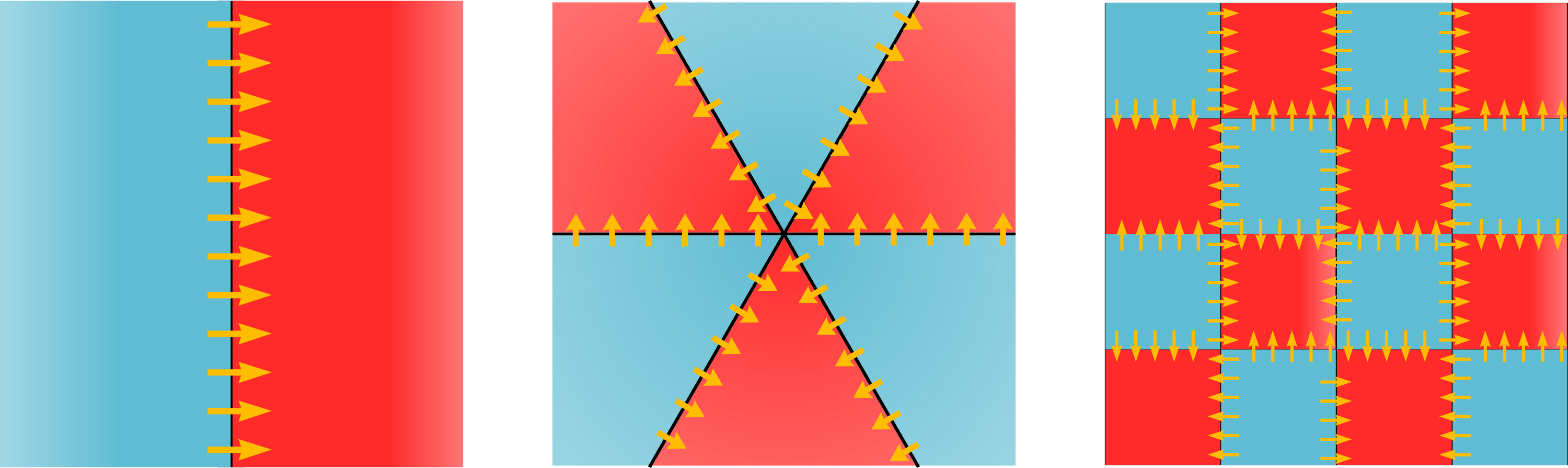}
\centering
\caption{Different samples of a surface divided into multiple magnetic regions. All these samples have infinitely many zero-modes since gradually adding dividing lines fits within the rectifying resolution.}
\label{fig:Samples}
\end{figure}

One can add any number of dividing lines in any orientation in order to enrich the simple example above. The dividing lines will break the surface into regions of upward and downward magnetic field. Let us define $\zeta (\vec{x}) \equiv \vec{B}\cdot \hat{\vec{z}}/|\vec{B}|$ to be one wherever the magnetic field is pointing up and minus one where it is pointing down while it is zero on the dividing lines. Then the zero mode solutions are obtained by replacing $\theta(\pm \vec{x})$ with $\theta_\pm(\vec{x})\equiv\frac{1}{2}(1\pm \zeta(\vec{x}))$ while the rectifying field, $\vec{S}(\vec{x}) \equiv \vect{\nabla}\zeta(\vec{x})$, is given by Dirac deltas sitting on the dividing lines and pointing away from downward magnetic regions. See Fig.~\ref{fig:Samples}. These configurations will always have zero modes.

We extend this solution to two additional examples: i)  a cylindrical topology $\mathbf{R}^1\times\mathbf{S}^1$ (with periodic vertical dividing lines separating alternating fields) and ii) a torus $\mathbf{S}^1\times\mathbf{S}^1$ (periodically tiled with square patches of alternating fields). We shall find that in contrast to the other examples (where we can always rectify the flat band) in the latter genuinely doubly periodic case, flat bands are only possible for certain ``magic'' values of the fields.

The cylindrical situation is provided by having $\vec{B}(x)=\zeta(x) B\hat{\vec{z}}$ with $\zeta(x)\equiv\sgn\left[\sin(\pi x/a)\right]$, which is a periodic step function with period $2a$, over which the magnetic field points along $\vec{\hat{z}}$ across $(0,a)$ and in the opposite direction across $(-a,0)$. We use the gauge freedom to choose the gauge field as $\vec{A}= B y \zeta(x)\hat{\vec{x}}$ noting that $\vect{\nabla}\cdot \vec{A}=\partial_x A_x$ is pure gauge. The normalizable zero mode solutions for this setup are given by,
\begin{equation}
    \Psi^c_k =  \left[ \begin{array}{cc}
      e^{\pm k(+ix +y)} e^{-\frac{B}{2}y^2} \theta_+(x) \\
        e^{\pm k(-ix +y)} e^{-\frac{B}{2}y^2} \theta_-(x)
    \end{array}  \right] \, ,
    \label{ZeroModeCyl}
\end{equation}
for each $k$, where as before the phase functions are (anti-)holomorphic functions $f^\pm$. Note the quasi-periodicity in $\hat{x}$ direction and also how our gauge choice is making normalizability of $\Psi^c_k$ manifest in contrast to a choice such as $\vec{A}=B x \zeta(x)\hat{\vec{y}}$. This implies that the cylindrical zero flux problem is inherently a two dimensional problem; it cannot be reduced to one dimension.

Because of the fact above we cannot simply identify $y=-b$ and $y=b$ and transition to a torus, since the transformation $y \rightarrow y + 2b$ leaves two terms behind in $e^{-By^2/2}\rightarrow e^{-By^2/2 - 2Bb^2 - 2Bby}$. Hence, the wave-function does not go back to itself unless these terms are canceled by extra terms granted by $f^\pm$. There is a unique class of functions $f^\pm$ that satisfies this condition.~\cite{TataTheta}

For the toroidal case, consider $\vec{B}(x,y)=\zeta(\vec{x})B \hat{\vec{z}}$ with $\zeta(\vec{x})= \zeta(x)\zeta(ay/b)$ generated by $\vec{A}=B y \zeta(\vec{x})\hat{\vec{x}}$. The zero mode solutions for this configuration are given by,
\begin{equation}
    \Psi^t_{\vec{k}} =  \left[ \begin{array}{cc}
      \vartheta_\vec{k}(x+iy) e^{-\frac{B}{2}y^2} \theta_+(\vec{x}) \\
       \vartheta_\vec{k}(x-iy) e^{-\frac{B}{2}y^2} \theta_-(\vec{x})
    \end{array}  \right] \, ,
    \label{ZeroModeTor}
\end{equation}
where $\vartheta_\vec{k}$ is a variation of the Jacobi theta function,
\begin{equation}
    \vartheta_\vec{k}(x\pm iy) \equiv \sum_{n\in \mathbb{Z}} e^{i\pi\left[i\frac{b}{a}\left(n + \frac{k_x a}{2\pi}\right)^2 + 2\left(n+ \frac{k_x a}{2\pi}\right)\left(\frac{x \pm iy}{a}- \frac{k_y b}{2\pi}\right)\right]} .
\end{equation}
The quasi-periodic behavior of $\vartheta_\vec{k}$ is given by $\vartheta_\vec{k} (x+a,y)= e^{ik_x a} \vartheta_\vec{k} (x,y)$ and,
\begin{equation}
    \vartheta_\vec{k} (x,y+2b)= e^{i2k_y b} e^{-i\frac{4\pi x}{a}} \vartheta_\vec{k} (x,y) e^{4\pi\frac{b}{a}  +\frac{4\pi}{a}y} \, .
\end{equation}
For the cancellation then it is required to have $4\pi b/a = 2Bb^2$ and $4\pi y/a = 2Bby$. This criterion is only satisfied for certain values of $B$, such that $ Bab = \int_\Box B dA = 2\pi$ with $\Box$ designating one magnetic square tile (note that the geometry of the tiles can be chosen differently, e.g., as triangles, $\tri$, in the moir{\'e} context~\cite{TopolCrit}).
If we replace $\vartheta_\vec{k}(x \pm iy)$ by $\vartheta^N_\vec{k}(x \pm iy)$ in our toroidal zero-mode, Eq.~\eqref{ZeroModeTor}, we arrive at another zero-mode. Therefore, the flat band criterion can be relaxed to
\begin{equation} \label{Crit}
    \int_\Box B dA = 2\pi N, \,\,\, \mbox{with } N \in \mathbb{Z} \, .
\end{equation}
Thus, the periodicity of the system is dictated by the dividing lines forming magnetic tiles and the criterion for localization is that the flux through each tile be an integer multiple of the flux quantum. This determines the ``magic'' values of the magnetic field, consistent with the topological criterion first argued in Ref.~\cite{TopolCrit}. 

The bilayer example of this \textit{magical} behaviour is obtained by following Eqs.~\eqref{SolBi}--\eqref{BiS} and it closely resembles that of the bilayer graphene. A similar result for moir\'e bilayer graphene has been demonstrated in Ref.~\cite{TopolCrit} where the appearance of flat bands was connected to chiral anomalies~\cite{Fujikawa,Fujikawa2004Book,ChiralAnomalyInt,ParhizkarPathInt} and the Atiyah-Singer index theorem~\cite{ASIndex,APSIndex}. The correspondence between the zero-flux localization picture presented here and the flat bands in moir\'e bilayer graphene is supported by three facts: First, that the general form of both theories is the same, see Eq.~\eqref{BiS}. Second, both constructions are generalizable to any lattice vector (c.f. Ref.~\cite{ParLocalizing}) as well as to a magnetic field that is inhomogeneous within a tile since the Poisson equation $\vec{\nabla}^2\phi=-B$ always has a solution, Eq.~\eqref{Poisson}.  Third, singular non-Abelian components mostly considered here can be smoothened (see Eq.~\eqref{Smeared} and the surrounding discussions). 

When the rectifying field, $\vec{S}$, coexists with the gauge field, $\vec{A}$, it effectively adds to the flux threading each title. This modifies the magic criterion, Eq.~\eqref{Crit}. This is exactly what happens in twisted bilayer graphene where without considering the influence of the $\vec{S}$ field the magic angle is given by $\alpha =1/\sqrt{3}\approx 0.577$ while reintroducing the effective flux yields $\alpha \approx 0.586$~\cite{TopolCrit,OriginVishwanath}. Through the process of \textit{Abelianizaion}~\cite{TopolCrit} (see also Ref.~\cite{NonAbelianBos}), one can push the $\vec{S}$ field to the boundaries or eliminate it, albeit at the cost of renormalizing the Abelian gauge field $\vec{A}$.

Note here that most results about zero flux localization are generalizable to the case of an electron with a quadratic dispersion, which can be obtained  by simply squaring  the Dirac Hamiltonian. In this case, by introducing a rectifying spin field near the regions where the magnetic field reversal occurs, the localization is restored. This is because $h$ and $h^2$ share zero modes: if $h\ket{0}=0$, $h^2\ket{0}=0$ as well.

Finally, we point out an exciting research program that the exact solutions found here offer in interacting systems. First, the presence of perfectly flat bands with a non-trivial spin texture suggests a new kind of fractional Hall states with zero total flux. Generalized Laughlin-like wave-functions can be written down using the spinors built of the elliptic functions that differ significantly from the textbook fractional Hall states. Second, as proposed in Ref.~\cite{ParLocalizing},  flat band localization may occur as an emergent phenomenon whereas almost dispersionless (heavy) electrons develop a self-consistent periodic texture - via a phase transition - where quasiparticle bands are flat. Apart from charge density and Cooper (superconducting) channels, this interaction-induced rectification of the nearly flat bands into an insulator may also occur in spin channels, depending on the energetics of the model. This general approach presents a highly attractive picture for a unifying theory of competing orders, and will be further developed elsewhere.


\acknowledgements
This work was supported by the National Science Foundation under Grant No. DMR-2037158, Army Research Office under Grant Number W911NF-23-1-0241, the Julian Schwinger Foundation and Simons Foundation.

\bibliographystyle{apsrev4-2}
\bibliography{main}

\end{document}